\documentclass[12pt,a4paper,final]{revtex4}

\usepackage{lmodern}
\usepackage{sidecap}
\usepackage{ulem}
\usepackage{epsfig}
\usepackage{amsmath,amssymb,amsthm}
\usepackage{graphicx}
\usepackage{bm}
\usepackage{color}

\usepackage{enumerate}

\usepackage{float}
\usepackage{tabularx}

\DeclareMathAlphabet{\mathpzc}{OT1}{pzc}{m}{it}

\begin{document}

\renewcommand{\textfraction}{0.00}


\newcommand{\vAi}{{\cal A}_{i_1\cdots i_n}}
\newcommand{\vAim}{{\cal A}_{i_1\cdots i_{n-1}}}
\newcommand{\vAbi}{\bar{\cal A}^{i_1\cdots i_n}}
\newcommand{\vAbim}{\bar{\cal A}^{i_1\cdots i_{n-1}}}
\newcommand{\htS}{\hat{S}}
\newcommand{\htR}{\hat{R}}
\newcommand{\htB}{\hat{B}}
\newcommand{\htD}{\hat{D}}
\newcommand{\htV}{\hat{V}}
\newcommand{\cT}{{\cal T}}
\newcommand{\cM}{{\cal M}}
\newcommand{\cMs}{{\cal M}^*}
\newcommand{\vk}{\vec{\mathbf{k}}}
\newcommand{\bk}{\bm{k}}
\newcommand{\kt}{\bm{k}_\perp}
\newcommand{\kp}{k_\perp}
\newcommand{\km}{k_\mathrm{max}}
\newcommand{\vl}{\vec{\mathbf{l}}}
\newcommand{\bl}{\bm{l}}
\newcommand{\bK}{\bm{K}}
\newcommand{\bb}{\bm{b}}
\newcommand{\qm}{q_\mathrm{max}}
\newcommand{\vp}{\vec{\mathbf{p}}}
\newcommand{\bp}{\bm{p}}
\newcommand{\vq}{\vec{\mathbf{q}}}
\newcommand{\bq}{\bm{q}}
\newcommand{\qt}{\bm{q}_\perp}
\newcommand{\qp}{q_\perp}
\newcommand{\bQ}{\bm{Q}}
\newcommand{\vx}{\vec{\mathbf{x}}}
\newcommand{\bx}{\bm{x}}
\newcommand{\tr}{{{\rm Tr\,}}}
\newcommand{\bc}{\textcolor{blue}}

\newcommand{\beq}{\begin{equation}}
\newcommand{\eeq}[1]{\label{#1} \end{equation}}
\newcommand{\ee}{\end{equation}}
\newcommand{\bea}{\begin{eqnarray}}
\newcommand{\eea}{\end{eqnarray}}
\newcommand{\beqar}{\begin{eqnarray}}
\newcommand{\eeqar}[1]{\label{#1}\end{eqnarray}}

\newcommand{\half}{{\textstyle\frac{1}{2}}}
\newcommand{\ben}{\begin{enumerate}}
\newcommand{\een}{\end{enumerate}}
\newcommand{\bit}{\begin{itemize}}
\newcommand{\eit}{\end{itemize}}
\newcommand{\ec}{\end{center}}
\newcommand{\bra}[1]{\langle {#1}|}
\newcommand{\ket}[1]{|{#1}\rangle}
\newcommand{\norm}[2]{\langle{#1}|{#2}\rangle}
\newcommand{\brac}[3]{\langle{#1}|{#2}|{#3}\rangle}
\newcommand{\hilb}{{\cal H}}
\newcommand{\pleft}{\stackrel{\leftarrow}{\partial}}
\newcommand{\pright}{\stackrel{\rightarrow}{\partial}}

\newcommand{\squeezeup}{\vspace{-2.5mm}}

\newcommand{\RomanNumeralCaps}[1]
    {\MakeUppercase{\romannumeral #1}}


\title{Extracting the temperature dependence in high-$p_\perp$ particle energy loss}

\date{\today}

\author{Stefan Stojku$^1$, Bojana Ilic$^1$, Marko Djordjevic$^2$ and Magdalena Djordjevic$^1$}

\affiliation{$^1$Institute of Physics Belgrade, University of Belgrade, Belgrade, Serbia \\
$^2$Faculty of Biology, University of Belgrade, Belgrade, Serbia}

\begin{abstract}
The suppression of high-$p_\perp$ particles is one of the main signatures of parton energy loss during its passing through the QGP medium, and is reasonably reproduced by different theoretical models. However, a decisive test of the reliability of a certain energy loss mechanism, apart from its path-length, is its temperature dependence. Despite its importance and
comprehensive dedicated studies, this issue is still awaiting for more stringent constraints. To this end, we here propose a novel observable to extract temperature dependence exponent of high-$p_{\perp}$ particle's energy loss, based on $R_{AA}$. More importantly, by combining analytical arguments, full-fledged numerical calculations and comparison with experimental data, we argue that this observable is highly suited for testing (and rejecting) the long-standing $\Delta E/E \propto L^2 T^3$ paradigm.
The anticipated significant reduction of experimental errors will allow direct extraction of temperature dependence, by considering different centrality pair in $A + A$ collisions (irrespective of the nucleus size) in high-$p_\perp$ region. Overall, our results imply that this observable,
which reflects the underlying energy loss mechanism, is very important to distinguish between different theoretical models.

\end{abstract}

\maketitle

\section{Introduction}
The main goal of ultra-relativistic heavy ion program~\citep{probe1,probe2,probe3,probe4} at RHIC and LHC is inferring the features of the created novel form of matter $-$ Quark-Gluon Plasma (QGP)~\citep{QGP1,QGP2}, which provides an insight into the nature of the hottest and densest known medium. Energy loss of rare high-$p_{\perp}$ partons traversing the medium is considered to be one of the crucial probes~\citep{Bj} of the medium properties, which also had a decisive role in QGP discovery~\citep{disc}. Comparison of predictions stemming from different energy loss models with experimental data,  tests our understanding of the mechanisms underlying the jet-medium interactions, thereby illuminating the QGP properties. Within this, an important goal presents a search for adequate observables for distinguishing the energy loss mechanisms.

Connected to this, it is well-known that the temperature ($T$) dependence of 
the energy loss predictions is directly related to the underlying energy loss 
mechanisms, e.g., pQCD {\it radiative} energy loss 
(BDMPS and ASW~\citep{BDMPS_parad,ASW_parad,paradigm2}; GLV~\citep{GLV0}; AMY~\citep{AMY1}; HT~\citep{HT0};
and some of their extensions~\citep{vP2,paradigm1,paradigm3,vP3,Majumd}) is 
typically considered to have cubic $T$ dependence ($T^3$, stemming from 
entropy, or energy density dependence), while {\it collisional} energy 
loss~\citep{Bj,TG,BT,HTcoll} is generally considered to be proportional to 
$T^2$. Additionally, AdS/CFT-motivated jet-energy loss 
models~\citep{AdS1,AdS2}  display even quartic ($T^4$) dependence on temperature. The different functional dependence on $T$ found in these models are the results of: considered energy loss mechanism (elastic or inelastic); different treatment of the QCD medium: finite or infinite size; inclusion or omission of finite temperature effects (i.e., application of temperature-modified, or vacuum-like propagators). 
 Therefore, assessing the accurate temperature dependence is important for disentangling relevant effects for adequate description of leading parton energy loss, and consequently for understanding the QGP properties.

For a comprehensive study on temperature (and path-length) dependence of different energy loss models we refer the reader to~\citep{vP2}. However, even this systematic study  couldn't single out $T$ dependence, as the attempt to simultaneously describe high-$p_\perp$ $R_{AA}$ and $v_2$ data  within these models requires some more rigorous physical justifications.
Moreover, the current error bars at the RHIC and the LHC are still too large to resolve between different energy loss models.
Having this in mind, we here propose a novel observable to extract the scaling of high-$p_{\perp}$ particle's energy loss on temperature. We expect that this observable will allow direct extraction of $T$ dependence from the data in the upcoming high-luminosity $3^{rd}$ run at the LHC, where the error-bars  are expected to notably decrease.

  We also propose high-$p_\perp$ $h^\pm$ as the most suitable probe for this study, as the experimental data for $h^\pm$ $R_{AA}$ are more abundant and with smaller error-bars, compared to heavier hadrons for all centrality classes, where this is also expected to hold in the future.
 Therefore, in this paper, we concentrate on $h^\pm$ in 5.02 TeV Pb + Pb collisions at the LHC, with the goal to elucidate this new observable, and test its robustness  to medium evolution
and colliding system size. By combining full-fledged numerical predictions and scaling arguments within our DREENA~\citep{DREENAC,DREENAB} framework, this  new observable yields the value of temperature dependence exponent, which is in accordance with our previous estimate~\citep{b}.  More importantly, we utilize this observable to question the long-standing $\Delta E/ E \propto L^2 T^3$ paradigm, used in a wide-range of theoretical models~\citep{BDMPS_parad,ASW_parad,vP2,paradigm1,paradigm2,paradigm3,vP3,GLV0,HT0,Majumd}.

\section{Theoretical framework}

In this study, we use our state-of-the-art dynamical energy loss formalism~\citep{DRad,DRad1,DColl}, which  includes several unique features in modeling jet-medium interactions: (1) The calculations within the finite temperature field theory and generalized Hard-Thermal-Loop approach~\citep{Kapusta}  (contrary to many models which apply vacuum-like propagators~\citep{BDMPS_parad,ASW_parad,GLV0,HT0}), so that infrared divergences are naturally regulated in a highly non-trivial manner; (2) Finite size of created QGP; (3) The QCD medium consisting of dynamical (moving) as opposed to static scattering centers, which allows the longitudinal momentum exchange with the medium constituents; (4) Both radiative~\citep{DRad,DRad1} and  collisional~\citep{DColl} contributions are calculated within the same theoretical framework; (5) The inclusion of finite parton's mass~\citep{masa}, making the formalism applicable to both  light and heavy flavor; (6) The generalization to a finite magnetic mass~\citep{Mmass}, running coupling~\citep{RunA} and beyond the soft-gluon approximation~\citep{bsga} is performed.

The analytical expression for single gluon radiation spectrum 
reads~\citep{DRad,Mmass,RunA,DREENAB}:
\begin{align}~\label{radEL}
\frac{dN_{rad}}{dx d \tau} {} & = \frac{C_2(G) C_R}{\pi}  \frac{1}{x}\int{\frac{d^2{\mathbf{q}}}{\pi}}
\frac{d^2{\mathbf{k}}}{\pi} \frac{\mu^2_E(T) - \mu^2_M(T)}{[{\mathbf{q}}^2 + \mu^2_E(T)] [{\mathbf{q}}^2 + \mu^2_M(T)]}  T \alpha_s(ET) \alpha_s\big(\frac{{\mathbf{k}}^2 + \chi(T)}{x} \big)  \nonumber \\
\times & \Big[1-\cos{\big( \frac{({\mathbf{k}} + {\mathbf{q}})^2 + \chi(T)}{xE^+} \tau \big)} \Big]
   \frac{2 ({\mathbf{k}}+{\mathbf{q}})}{({\mathbf{k}}+{\mathbf{q}})^2 + \chi(T)} \Big[\frac{{\mathbf{k}}+{\mathbf{q}}}{({\mathbf{k}}+{\mathbf{q}})^2 + \chi(T)} - \frac{{\mathbf{k}}}{{\mathbf{k}}^2 + \chi(T)} \Big] ,
\end{align}
where ${\mathbf{k}}$ and ${\mathbf{q}}$ denote transverse momenta of radiated and exchanged gluons, respectively,  $C_2(G) = 3$, $C_R=4/3$ ($C_R=3$) for quark (gluon) jet, while $\mu_E(T)$ and $\mu_M(T)$ are electric (Debye) and magnetic screening masses, respectively. Temperature dependent Debye mass~\citep{DREENAB,Deb} is obtained by self-consistently solving Eq.~(5) from Ref.~\citep{DREENAB}. $\alpha_s$ is the (temperature dependent) running coupling~\citep{Field,RunA,DREENAB}, $E$ is the initial jet energy, while $\chi(T) =  M^2x^2 + m^2_g(T)$, where $x$ is the longitudinal momentum fraction of the initial parton carried away by the emitted gluon. $M$ is the mass of the propagating parton, while the gluon mass is considered to be equal to its asymptotical mass $m_g = \mu_E/ \sqrt{2}$~\citep{mg}.

 The analytical expression for collisional energy loss per unit length is given by the following expression~\citep{DColl,DREENAB}:
 \begin{align}~\label{collEL}
 & \frac{dE_{coll}}{d \tau} =  \frac{2 C_R}{\pi v^2} \alpha_s(ET) \alpha_s(\mu^2_E(T)){\int_0^{\infty}} n_{eq}(|\vec{{\mathbf{k}}}|,T)d |\vec{{\mathbf{k}}}| \nonumber \\
 \times & \Big[\int_0^{|{\vec{{\mathbf{k}}}}|/(1+v)} d |\vec{{\mathbf{q}}}| \int_{-v |\vec{{\mathbf{q}}}|}^{v |\vec{{\mathbf{q}}}|} \omega d \omega  + \int_{|{\vec{{\mathbf{k}}}}|/(1+v)}^{|\vec{{\mathbf{q}}}|_{max}} d |\vec{{\mathbf{q}}}|  \int_{ |\vec{{\mathbf{q}}}| -2|\vec{{\mathbf{k}}}| }^{v |\vec{{\mathbf{q}}}|} \omega d \omega \Big]  \\
\times & \Big[|\Delta_L(q,T)|^2 \frac{(2 |\vec{{\mathbf{k}}}| + \omega)^2 - |\vec{{\mathbf{q}}}|^2}{2} +  |\Delta_T(q,T)|^2 \frac{(|\vec{{\mathbf{q}}}|^2 - \omega^2) ((2 |\vec{{\mathbf{k}}}| + \omega)^2 + |\vec{{\mathbf{q}}}|^2)}{4 |\vec{{\mathbf{q}}}|^4} (v^2 |\vec{{\mathbf{q}}}|^2 - \omega^2)\Big], \nonumber
\end{align}
where $n_{eq}(|\vec{{\mathbf{k}}}|, T) = \frac{N}{e^{|\vec{{\mathbf{k}}}|/T} -1} + \frac{N_f}{e^{|\vec{{\mathbf{k}}}|/T} + 1}$ is the equilibrium momentum distribution ~\citep{BT} including gluons, quarks and antiquarks.
 $k$ is the 4-momentum of the incoming medium parton, $v$ is velocity of the initial jet and $q = (\omega, \vec{{\mathbf{q}}})$ is the 4-momentum of the exchanged gluon.  $|\vec{{\mathbf{q}}}|_{max}$ is provided in Ref.~\citep{DColl}, while $\Delta_T (q,T)$ and  $\Delta_L(q,T)$ are effective transverse and longitudinal gluon propagators given by Eqs. (3) and (4) from Ref.~\citep{DREENAB}.

 Despite very complicated temperature dependence of Eqs.~(\ref{radEL}) and (\ref{collEL}), in~\citep{b} it was obtained that our dynamical energy loss formalism~\citep{RunA} (which accommodates some of unique jet-medium effects mentioned above) has an exceptional feature of near linear $T$ dependence. That is, while $T^3$ dependence for radiative energy loss is widely used~\citep{BDMPS_parad,ASW_parad,paradigm2,GLV0,AMY1,HT0,paradigm3,paradigm1,vP2,vP3,Majumd}, from Eq.~\eqref{radEL} it is evident that this simplified relation is reproduced with approximations of using vacuum gluon propagators (leading to the absence of $m_g(T)$ from $\chi$ expression) and neglecting running coupling. It is straightforward to show that in that case leading $T$ dependence is: $\frac{\Delta E_{rad}}{E} \propto \mu^2_E T \propto T^3$ ($\mu_E \propto T$). However, Eq.~\eqref{radEL} clearly demonstrates that a more realistic $T$ dependence is far from cubic, where in~\citep{b} it was shown that asymptotic $T$ dependence of our full radiative energy loss is between linear and quadratic.

 Additionally, commonly overlooked (due to being smaller compared to radiative at high-$p_\perp$) collisional energy loss, must not be neglected in suppression predictions~\citep{ELeffects}. Moreover, widely-used dominant $T^2$ dependence of collisional energy loss~\citep{Bj,BT,TG,HTcoll} can also be shown to be a consequence of: {\it i)} using tree-level diagrams, and consequently introducing artificial cut-offs to non-physically regulate ultraviolet (and infrared) divergencies (e.g., in~\citep{Bj}) in the hard  momentum transfer sector~\citep{BT}; or  {\it ii)} considering only soft momentum exchange~\citep{TG}. 
That is, it is straightforward to show that Eq.~(\ref{collEL}) recovers leading $T^2$ dependence from~\citep{TG} if: 1) only soft gluon sector is considered, with upper limit of integration  artificially set to $|\vec{{\mathbf{q}}}|_{max}$; 2) only forward emission is accounted for ($\omega > 0$); and 3) running coupling is neglected. Accordingly, in~\citep{b} it was demonstrated that complex $T$ dependence of our {\it collisional energy loss} (Eq.~\eqref{collEL}) reduces not to commonly considered quadratic, but rather nearly linear dependence for asymptotically large $p_\perp$. Therefore, a state-of-the-art energy loss model leads to a much slower growth of the energy loss with temperature compared to common paradigm, where the widely assumed faster growth can be reproduced only through quite drastic simplifying assumptions.

Since the goal of this paper is the extraction of the temperature dependence exponent of the energy loss, this study will furthermore provide an opportunity to test our dynamical energy loss formalism on more basic level.

\section{Numerical framework}

In this paper, the predictions are generated within our fully optimized DREENA~\citep{DREENAC,DREENAB} numerical framework, comprising: {\it i)} Initial parton momentum distribution ~\cite{ID}; {\it ii)} Energy loss probability based on
our dynamical energy loss formalism~\citep{DColl,DRad,DRad1} (discussed in the previous section),
which includes multi-gluon~\cite{MGF} and path-length fluctuations~\cite{PLF}. The path-length fluctuations are calculated according to the procedure provided in~\citep{Dainese},
(see also~\citep{DREENAC}); and {\it iii)}  Fragmentation functions~\citep{FF}.

In generating numerical predictions for comparison with 5.02 TeV Pb + Pb  data for different centrality classes, we use {\it no fitting parameters}, i.e., all the parameters correspond to standard literature values, and for their values we refer reader to~\citep{DREENAC}.

In the first part of our study, the average temperature  for each centrality class is obtained according to the procedure outlined in Refs.~\citep{NonC,DREENAC}. Similarly, initial temperature ($T_0$) for each centrality, in a part of this study where we test the sensitivity of the obtained conclusions to the medium evolution, is estimated in accordance with~\citep{DREENAB}.

\section{Results and discussion}

In this section, we first address the choice of the
 suitable observable for extracting energy loss temperature dependence. For this purpose, an observable which is
 sensitive only to the details of jet-medium interactions (to facilitate extraction of $T$ dependence),
 rather than the subtleties of medium evolution (to avoid unnecessary complications
 and ensure robustness), would be optimal. $R_{AA}$ has such features, since it was previously reported that it is very  sensitive to energy loss effects~\citep{ELeffects} and the average medium properties, i.e., average temperature, while being practically insensitive to the details of medium evolution (in distinction to $v_2$)~\citep{Renk,Molnar,DREENAB,DREENAC,NewObserv}. Therefore, it is plausible that the appropriate observable should be closely related to $R_{AA}$.

Our theoretical and numerical approaches described above (where the dynamical energy loss explicitly depends on $T$), are implemented in a fully optimized DREENA framework~\citep{DREENAC,DREENAB}, which makes it suitable for this study. To more easily interpret the obtained results, we start from constant $T$  medium, i.e., DREENA-C~\citep{DREENAC} and continue toward evolving medium case, i.e., DREENA-B  framework~\citep{DREENAB}. We here exploit that DREENA-C and DREENA-B are analytically trackable, allowing to derive appropriate scaling behavior. To additionally test the obtained results, we will then use our DREENA-A (A stands for Adaptive) framework, which employs full 3+1D hydrodynamics evolution~\cite{Molnar:2014zha}.

With the intention of extracting simple functional dependence on $T$ (of the otherwise analytically and numerically quite complex dependence of the fractional energy loss, see Eqs.~(\ref{radEL},~\ref{collEL})), we first provide the scaling arguments. These scaling (analytical) arguments will then be followed by a full-fledged numerical analysis. Namely, in~\citep{MGF, DREENAC,DREENAB,NewObserv} it was shown that, at very large values of transverse momentum $p_{\perp}$  and/or in peripheral collisions, the following estimates can be made:
\begin{align}
    \Delta E / E & \approx  \eta {T}^a {L}^b, \nonumber \\
    R_{AA} &  \approx 1-\xi {T}^a {L}^b,
    \label{fractEloss}
\end{align}
where $\eta$ denotes a proportionality factor, depending on initial parton transverse momentum and its flavor, while  $\xi = (n-2)\eta / 2$, where $n$ is the steepness of a power law fit to the initial transverse momentum distribution, i.e., $d\sigma/dp^2_{\perp} \propto p^{-n}_\perp$. ${T}$ and ${L}$ denote the average temperature (of the QCD medium) along the jet path and the average path length traversed by the energetic parton. The scaling factors for temperature and path-length energy loss dependence are denoted as $a$ and $b$, respectively.

We next formulate the following quantity $R^T_{AA}$, with the goal to isolate the temperature dependence:
\begin{align}
    R^T_{AA} =\frac{1-R_{AA}}{1-R^{ref}_{AA}},
    \label{ratioIntroduction0}
\end{align}
which presents $(1-R_{AA})$ ratio for a pair of two different centrality classes. The centrality class that corresponds to $R^{ref}_{AA}$ (i.e., the quantity in the denominator) is denoted as the referent centrality, and is  always lower (corresponding to more central collision) than centrality in the numerator. We term this new quantity, given by Eq.~\eqref{ratioIntroduction0}, as a {\it temperature dependent suppression ratio} ($R^T_{AA}$), which we will further elucidate below.

Namely, by using Eq.~\eqref{fractEloss}, it is straightforward to isolate average $T$ and average path-length
\begin{figure}
  \includegraphics[width=0.4\linewidth]{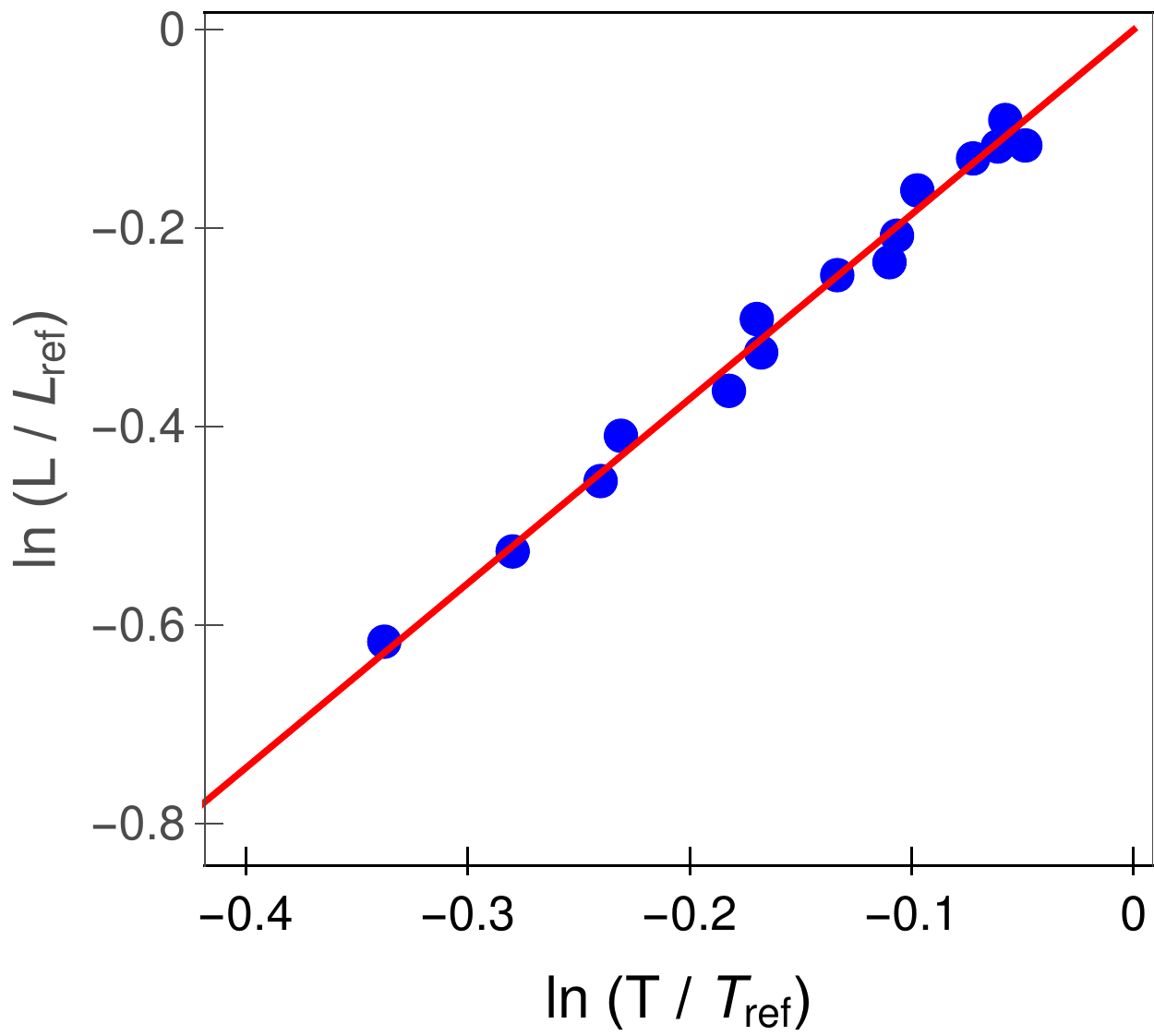}
  \caption{$\boldsymbol{\ln({L}/ {L}_{ref})}$ {\textbf{\textit{vs.}}} $\boldsymbol{\ln ({T}/{T}_{ref})}$ in 5.02 TeV Pb+Pb collisions at the LHC for various centrality pairs. The {\it referent centralities} (for quantities in denominators) acquire one of the following values: ($5-10 \%, 10-20\%, 20-30 \%, 30-40\%, 40-50\%$); while the centralities in the numerator are always higher (the highest one being $50 - 60 \%$).
  The solid red line corresponds to the linear fit to the calculated points.}
  \label{fig:sl1}
\end{figure}
 dependence of $R^T_{AA}$:
\begin{align}
    R^T_{AA} =\frac{1-R_{AA}}{1-R^{ref}_{AA}} &\approx \frac{\xi {T}^a {L}^b}{\xi {T}_{ref}^a {L}_{ref}^b } = \bigg( \frac{{T}}{{T}_{ref}} \bigg) ^a \cdot \bigg( \frac{{L}}{{L}_{ref}} \bigg) ^b,
    \label{ratioIntroduction}
\end{align}
which in logarithmic form reads:
\begin{eqnarray}
 \ln(R^T_{AA}) =   \ln \bigg( \frac{1-R_{AA}}{1-R^{ref}_{AA}} \bigg) \approx a \ln \bigg( \frac{{T}}{{T}{ref}} \bigg) + b \ln \bigg( \frac{{L}}{{L}_{ref}} \bigg).
    \label{lnIntroduction}
\end{eqnarray}

However, the remaining dependence of the newly defined quantity on the path length is undesired for the purpose of this study. So, in order to make use of the previous equation,  we first test how the two terms on the right-hand side of Eq.~\eqref{lnIntroduction} are related. To this end, in Fig.~\ref{fig:sl1} we plot $\ln({L} /{L}_{ref})$ against $\ln({T} /{T}_{ref})$ for several combinations of centralities, as denoted in the caption of Fig.~\ref{fig:sl1}. For a particular centrality class, ${L}$ and ${T}$ are calculated by averaging over path-length distributions given in~\citep{DREENAC}.

Conveniently, Fig.~\ref{fig:sl1} shows a linear dependence $ \ln({L} /{L}_{ref})\approx k \ln({T} /{T}_{ref})$, with $k \approx 1.86$.
 This leads to a simple relation:
\begin{align}
  \ln(R^T_{AA}) & \approx (a+kb) \ln \bigg( \frac{{T}}{{T}_{ref}} \bigg), \label{lnOnlyT1}
\end{align}
so that with $f=a+k b$:
\begin{align}
  R^T_{AA} & \approx \Big( \frac{{T}}{{T}_{ref}} \Big)^{f},
    \label{lnOnlyT}
\end{align}
where this simple form facilitates extraction of $a$.
\begin{figure}
  \includegraphics[width=\linewidth]{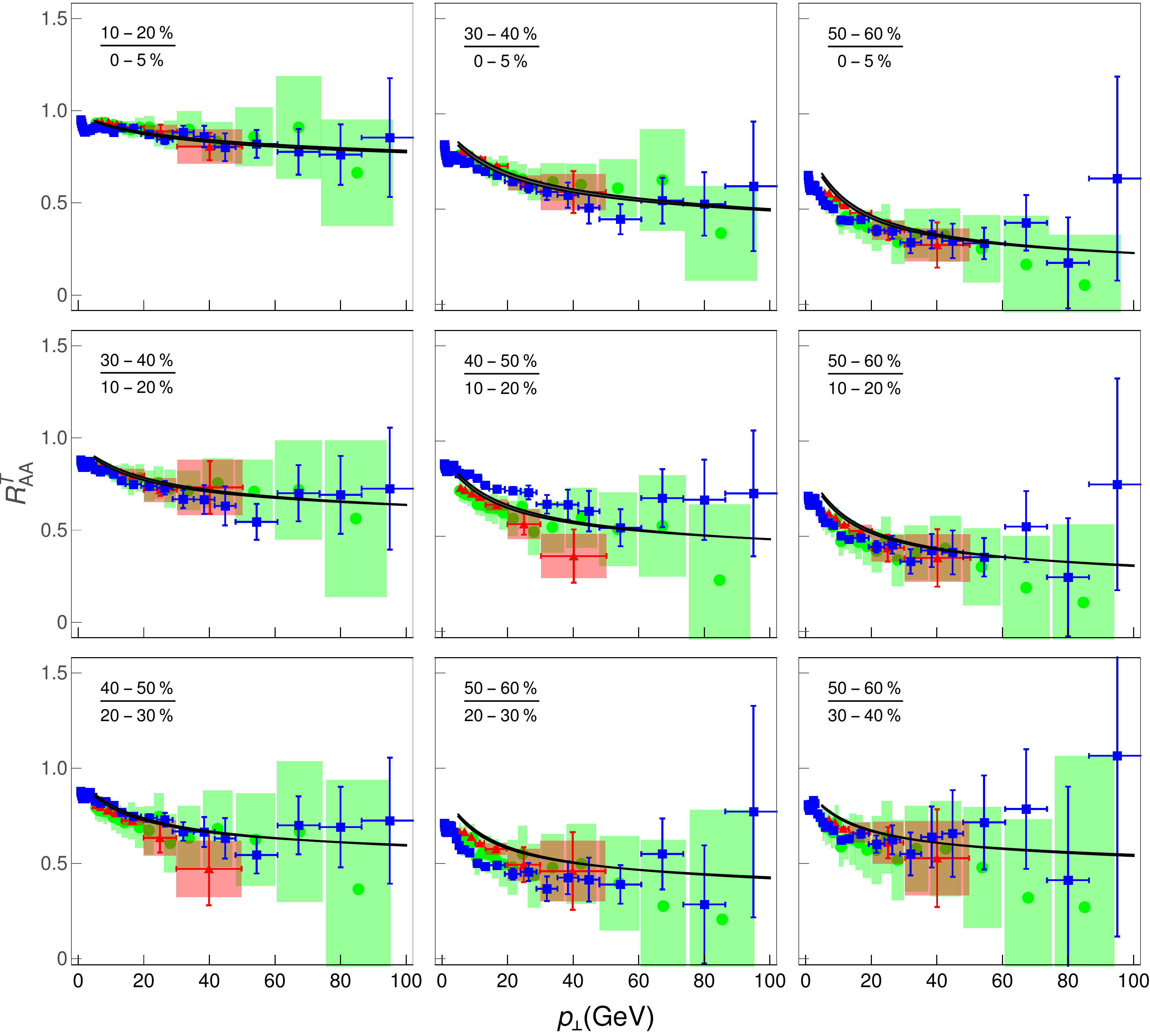}
  \caption{{\bf{Charged hadron  $\boldsymbol{R^T_{AA}}$ for different pairs of centrality classes   as a function of $p_\perp$.}}
   The predictions generated within our full-fledged suppression numerical procedure DREENA-C~\citep{DREENAC} (black curves with corresponding gray bands) are compared with ALICE~\citep{ALICE} (red triangles), CMS~\citep{CMS} (blue squares) and ATLAS~\citep{ATLAS} (green circles) data. The lower (upper) boundary of each band corresponds to $\mu_M / \mu_E = 0.6$ ($\mu_M / \mu_E = 0.4$). Centrality pairs are  indicated in the upper-left corner of each plot.}
  \label{fig:sl2}
\end{figure}

In Eq.~\eqref{lnOnlyT}, $R^T_{AA}$ depends solely on ${T}$ and effectively temperature dependence exponent $a$ (as $k$ and $b$~\citep{NewObserv} are known), which justifies the use of "temperature-sensitive" term with this new quantity. Therefore, here we propose $R^T_{AA}$, given by Eq.~\eqref{ratioIntroduction0}, as a new observable,
which is highly suitable for the purpose of this study.

The proposed extraction method is the following: We use our full-fledged DREENA-C numerical procedure to generate predictions for $R_{AA}$ and thereby for the left-hand side of Eq.~\eqref{lnOnlyT}. Calculation of average ${T}$ is already outlined in the previous section.
We will generate the predictions with full-fledged procedure, where we expect asymptotic scaling behavior (given by Eq.~\eqref{lnOnlyT}) to be valid at high $p_{\perp}\approx 100$ GeV. Having in mind that values of $k$ and $b$ parameters have been extracted earlier, the temperature dependence exponent $a$ in very high-$p_{\perp}$ limit can then be estimated from slope ($f$) of a $\ln(R^T_{AA})$ {\it vs.} $\ln({T}/{T}_{ref})$ linear fit, done for a variety of centrality pairs.

However, before embarking on this task, we first verify whether our predictions of $R^T_{AA}$ for different centrality classes, based on the full-fledged DREENA-C framework, are consistent with the available experimental data. In Fig.~\ref{fig:sl2} we compare our $R^T_{AA}$ {\it vs.} $p_\perp$ predictions for charged hadrons with corresponding 5.02 TeV Pb + Pb LHC data from ALICE~\citep{ALICE}, CMS~\citep{CMS} and ATLAS~\citep{ATLAS}, for different centrality pairs as indicated in the upper-left corner of each plot. Despite the large error bars, for all centrality pairs we observe consistency between our DREENA-C predictions and experimental data, in $p_\perp$ region where our formalism is applicable ($p_{\perp}\gtrsim 10$ GeV). Moreover, we also notice the flattening of each curve with increasing $p_\perp$ ($\sim 100$) GeV, confirming that the expecting saturating (limiting) behavior is reached.
\begin{figure}
  \includegraphics[width=0.4\linewidth]{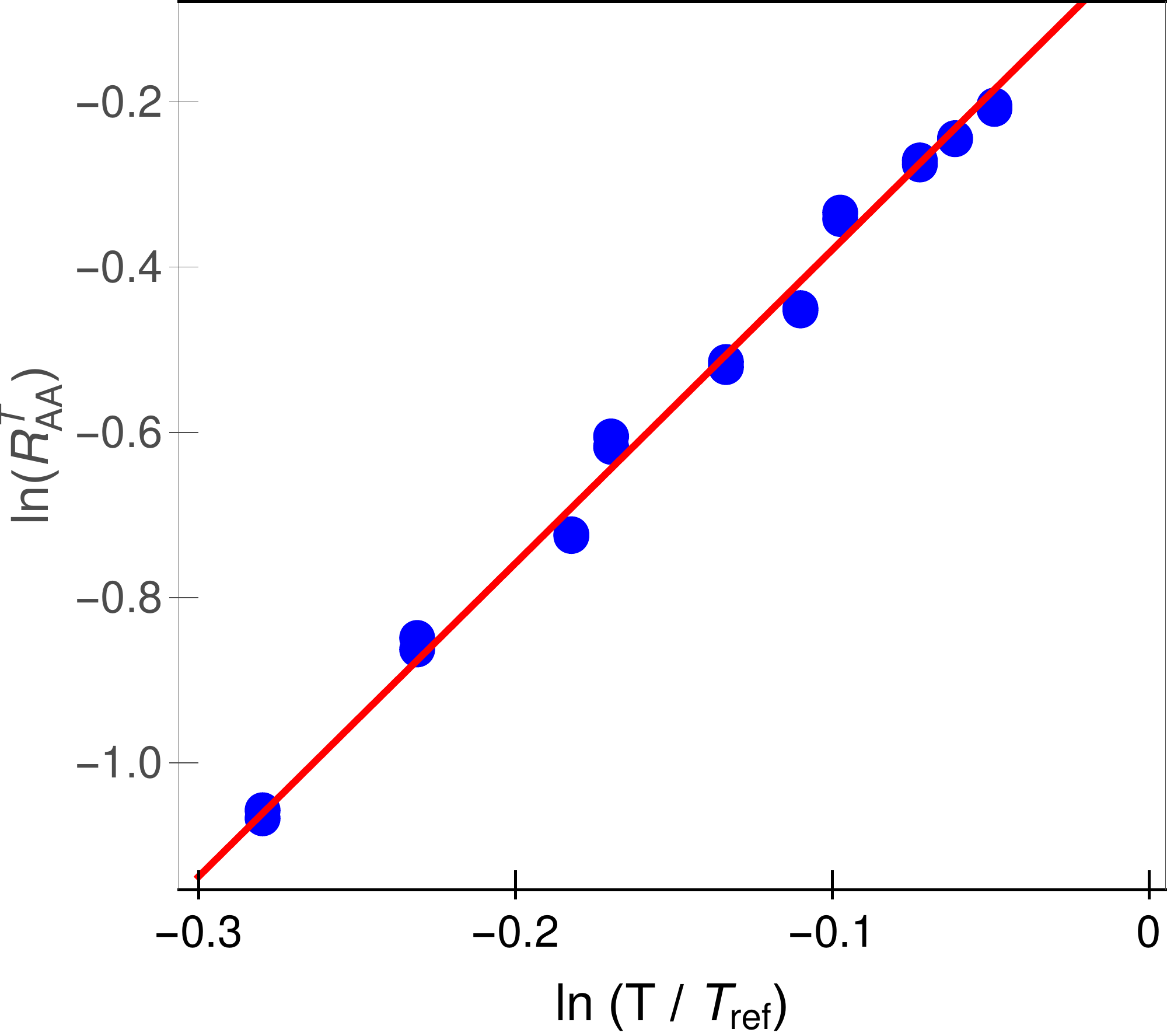}
  \caption{$\boldsymbol{\ln(R^T_{AA})}$ {\textbf{\textit{ vs.}}} $\boldsymbol{\ln({T}/{T}_{ref})}$ {\bf relation.} $\ln(R^T_{AA})$ and  $\ln({T}/{T}_{ref})$ are calculated from full-fledged DREENA-C framework~\citep{DREENAC}, for $h^{\pm}$ at $p_{\perp} = 100$ GeV in 5.02 TeV Pb+Pb collisions at the LHC for different centrality pairs. The referent centrality values are: ($10-20 \%, 20-30  \%, 30-40  \%, 40-50  \%$), while their counterpart values are always higher, with highest being equal to $50-60 \%$.  The red solid line corresponds to the linear fit to the values. Remaining parameters are the same as in Fig.~\ref{fig:sl2}.}
  \label{fig:sl4}
\end{figure}

Furthermore, based on the analytical relation provided by Eq.~\eqref{lnOnlyT1}, we expect linear functional dependence between $\ln{R^T_{AA}}$ and $\ln({T}/{T_{ref}})$, which we test in Fig.~\ref{fig:sl4}. Note that all quantities throughout the paper are determined at $p_\perp=100$ GeV, and by calculating $R^T_{AA}$  for various centrality pairs (see figure captions) within full-fledged DREENA procedure. Remarkably, from Fig.~\ref{fig:sl4}, we observe that $\ln(R^T_{AA})$ and $\ln(T/T_{ref})$ are indeed linearly related, which confirms the validity of our scaling arguments at high-$p_{\perp}$ and the proposed procedure.

 Linear fit to calculated points in Fig.~\ref{fig:sl4} leads to the proportionality factor $f = a + kb = 3.79 \sim 4$. This small value of $f$ would lead to $k$ smaller than $1$ {\it if} (commonly assumed) $a=3$ and $b=2$ are used. Such $k$ value seems however implausible, as it would require $(T/T_{ref})$ to change more slowly with centrality compared to $(L/L_{ref})$.

 More importantly, the temperature exponent can now be extracted ($b \approx 1.4$ as estimated in~\citep{NewObserv}), leading to $a \approx 1.2$. This indicates  that temperature dependence of energetic particle energy loss (at very high $p_\perp$) is close to linear (see Eq.~\eqref{fractEloss}), that is,  certainly not quadratic or cubic, as commonly considered. This is in accordance with previously reported dependence of fractional dynamical energy loss on ${T}$ to be somewhere between linear and quadratic~\citep{b}, and as opposed to commonly used pQCD estimate  $a=3$ for radiative~\citep{BDMPS_parad,ASW_parad,vP2,paradigm1,paradigm2,AMY1,paradigm3,vP3,GLV0,HT0,Majumd} (or even $a=2$ for collisional~\citep{Bj,TG,BT,HTcoll}) energy loss.

\begin{figure}
  \includegraphics[width=0.8\linewidth]{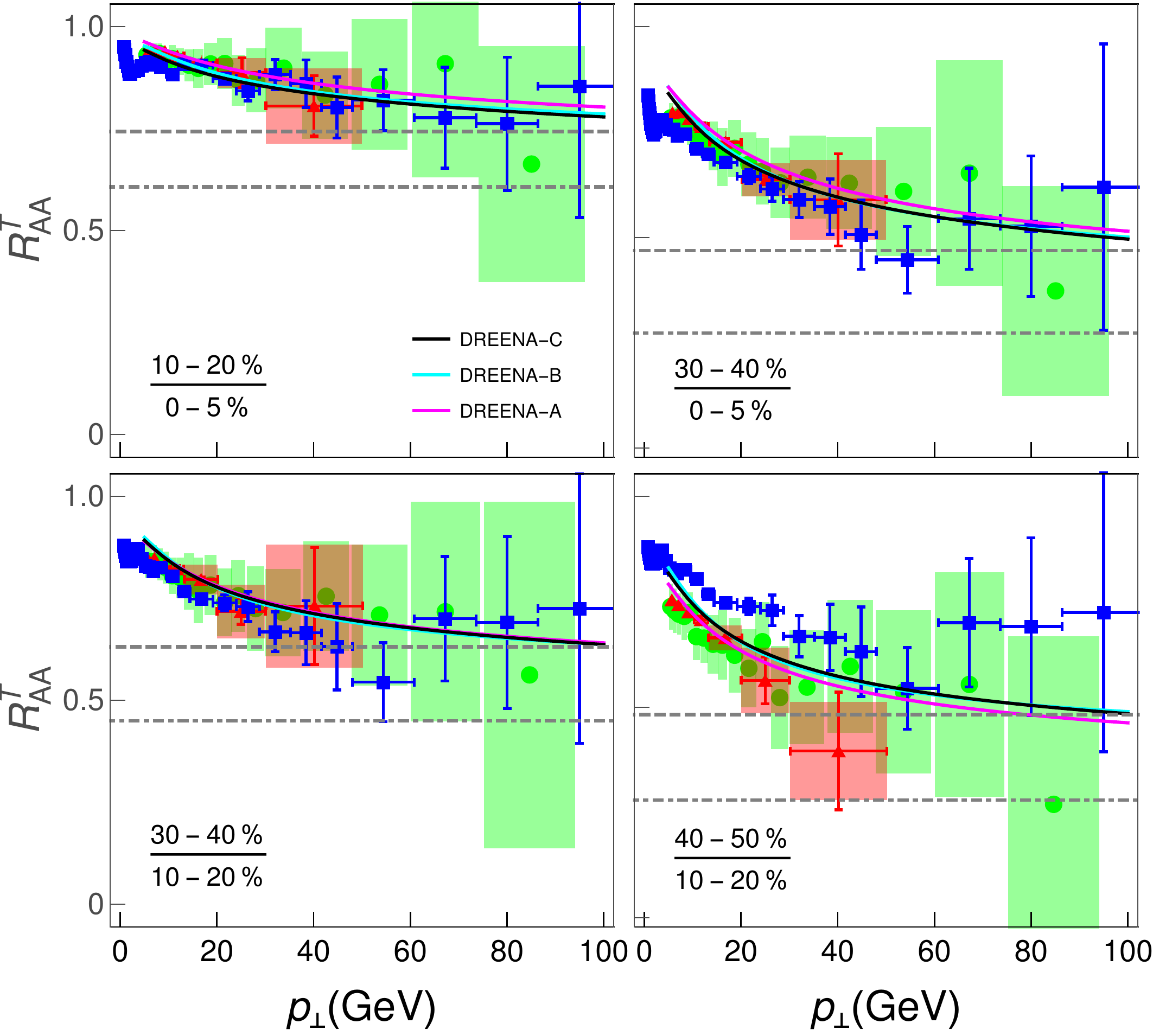}
 \caption{{\bf{The discriminative power of $\boldsymbol{R^T_{AA}}$ quantity in resolving energy loss mechanism.}} Four panels in Fig.~\ref{fig:sl2} are extended to include comparison of our asymptotic scaling behavior $({T}/{T}_{ref})^{1.2} \cdot ({L}/{L}_{ref})^{1.4}$ (gray dashed horizontal line) with common assumption $(T/T_{ref})^{3} \cdot (L/L_{ref})^{2}$  (gray dot-dashed horizontal line). The figure also shows comparison of $R_{AA}^T$s obtained by three different numerical frameworks: constant temperature DREENA-C (black curve), 1D Bjorken expansion DREENA-B~\citep{DREENAB} (cyan curve) and full 3+1D hydrodynamics evolution~\cite{Molnar:2014zha} DREENA-A (magenta curve). The remaining labeling is the same as in Fig.~\ref{fig:sl2}.}
  \label{fig:sl}
\end{figure}

The extraction of $T$ dependence, together with previously estimated path-length dependence~\citep{NewObserv}, within DREENA framework, allows utilizing this new observable $R^T_{AA}$ in discriminating between energy loss models, with the aim of better understanding QGP properties. To this end, in Fig.~\ref{fig:sl}, we {\it i}) Test sensitivity of $R^T_{AA}$ on different medium evolutions (constant temperature, 1D Bjorken~\cite{Bjorken} and full 3+1D hydrodynamics~\cite{Molnar:2014zha}). {\it ii}) Compare the asymptote derived from this study ($(T/T_{ref})^{1.2} \cdot (L/L_{ref})^{1.4}$), with commonly used estimate of $(T/T_{ref})^{3} \cdot (L/L_{ref})^{2}$.

Several conclusions can be drawn from Fig.~\ref{fig:sl}: {\it i}) With respect to different models of QGP expansion, we see that, as expected, obtained  $R^T_{AA}$ results are similar, i.e., not very sensitive to the details of the medium evolution. As in DREENA-C (and DREENA-B, see the next subsection) the temperature dependence can be analytically tracked (which is however not possible in more complex DREENA-A), this result additionally confirms that DREENA-C framework is suitable for the extraction of energy loss temperature dependence. {\it ii}) Ideally, $T$ dependence exponent could be directly extracted from experimental data, by fitting a straight line to very high-$p_\perp$ part ($\sim 100$ GeV) of $R^T_{AA}$ for practically any centrality pair. However, the fact that data from different experiments (ALICE, CMS and ATLAS) are not ideally consistent, and that the error-bars are quite sizeable, currently prevents from such direct extraction. The error-bars in the upcoming high-luminosity $3^{rd}$ run at the LHC are however expected to significantly decrease, which would enable the direct extraction of the exponent $a$ from the data. {\it iii)} From Fig.~\ref{fig:sl} it importantly follows that even the current large experimental uncertainties seem to indicate inadequacy of commonly exploited energy loss dependence $\propto T^3 L^2$, practically for all considered centrality pairs. This is consistent with our other results, which all indicate breaking of the long-standing paradigm of energy loss temperature and path-length  dependence. Future increase in measurements precision could provide the confidence to this conclusion and resolve the exact form of these dependencies from the data, through our proposed observable. This discriminative power of $R^T_{AA}$ quantity highlights its importance in understanding the underlying energy loss mechanisms in QGP.

\subsection{Effects of medium evolution}
\begin{figure}
  \includegraphics[width=0.4\linewidth]{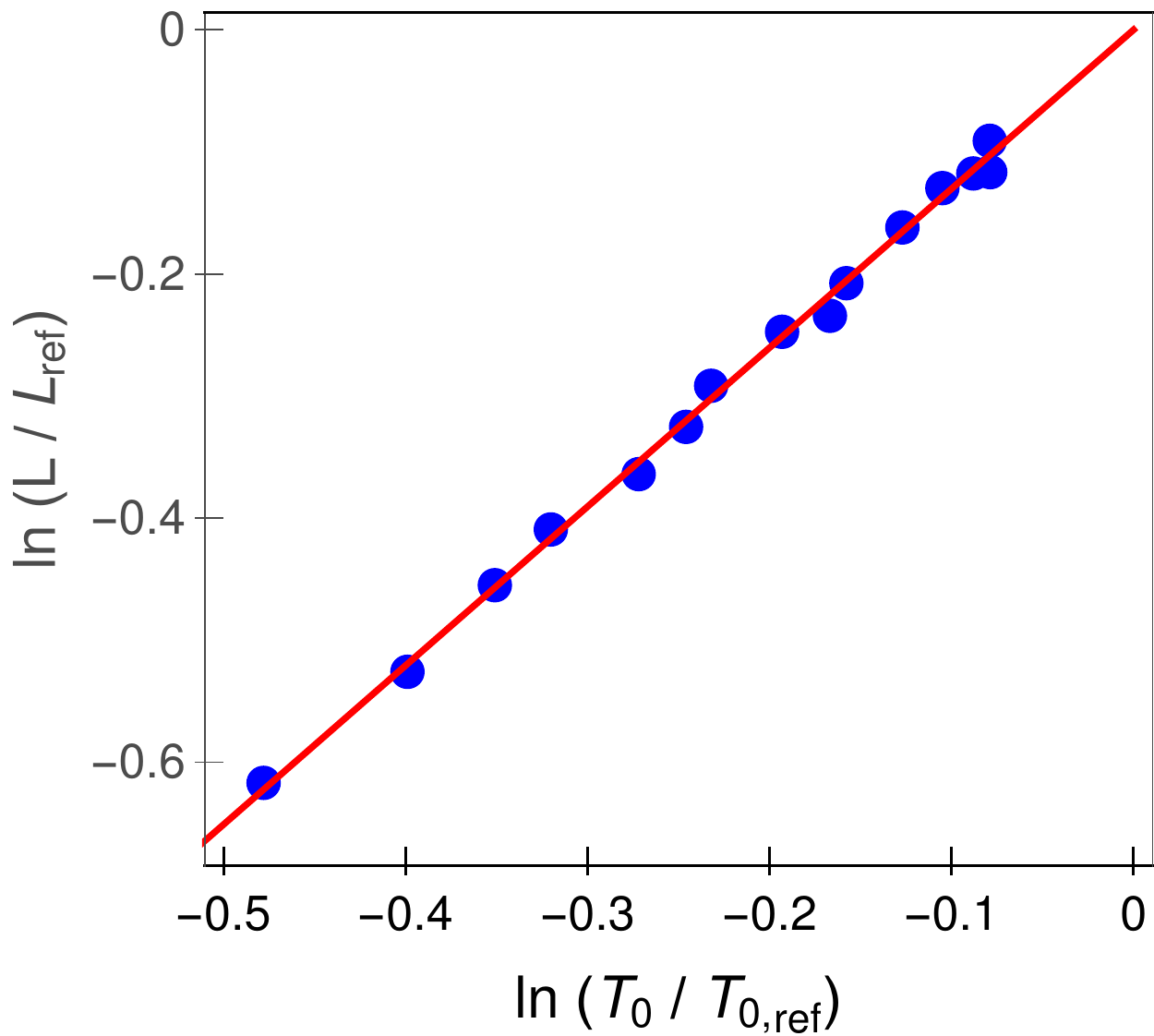}
  \caption{{$\boldsymbol{\ln(L/L_{ref})}$ {\textbf{\textit{vs.}}} $\boldsymbol{\ln(T_0/T_{0,ref})}$ {\bf for various pairs of centralities in evolving medium.}} The assumed centrality pairs are the same as in Fig.~\ref{fig:sl1}. The red solid line corresponds to the linear fit to the values.}
  \label{fig:sl7}
\end{figure}

While in Fig.~\ref{fig:sl} we showed that $R^T_{AA}$ results are robust with respect to the medium evolution, the analytical procedure for  extracting temperature dependence is different in DREENA-C and DREENA-B frameworks. Comparing scaling factors extracted from these two procedure, can be used to test reliability of the proposed procedure. In this subsection, we consequently utilize DREENA-B framework~\citep{DREENAB}, where medium evolution is introduced through Bjorken 1D hydrodynamical expansion~\citep{Bjorken}, i.e., there is the following functional dependence of $T$ on path-length:
\begin{equation}
    T = T_0 \cdot \Big( \frac{\tau_0}{l} \Big) ^ {1/3},
    \label{BjorkenMain}
\end{equation}
where $T_0$ and $\tau_0 = 0.6$ fm~\citep{tau0KH,tau0BMB} denote initial temperature and thermalization time of the QGP.  Proceeding in the similar manner as in constant medium case, $R^T_{AA}$ (given by Eq.~\eqref{ratioIntroduction0}) in evolving medium (for coupled $T$ and $l$, where $l$ stands for traversed path length) reads:
\begin{align}
  R^T_{AA} &= \frac{\int_{0}^{L} T^a l^{b-1} dl}{\int_{0}^{L_{ref}} (T_{ref})^a (l_{ref})^{b-1} dl_{ref}} = \frac{T^a_0 \tau^{a/3}_0 \int_{0}^{L} \frac{l^{b-1}}{l^{a/3}} dl}{T^a_{0,ref} \tau^{a/3}_0 \int_{0}^{L_{ref}} \frac{(l_{ref})^{b-1}}{(l_{ref})^{a/3}} dl_{ref}}
    = \bigg( \frac{T_0}{T_{0,ref}} \bigg)^a \cdot \bigg( \frac{L}{L_{ref}} \bigg) ^{b-\frac{a}{3}},
    \label{RatioBjorken1}
\end{align}
where we used Eq.~\eqref{BjorkenMain}. Again, we assess whether there is a simple relation between logarithms of (now initial) temperature ratio and average path-length ratio for different centrality pairs. Similarly to constant $T$ case, from Fig.~\ref{fig:sl7} we infer linear dependence between these two quantities, where slope coefficient now acquires the value $\kappa \approx 1.3$. Thus,
\begin{figure}
  \includegraphics[width=0.4\linewidth]{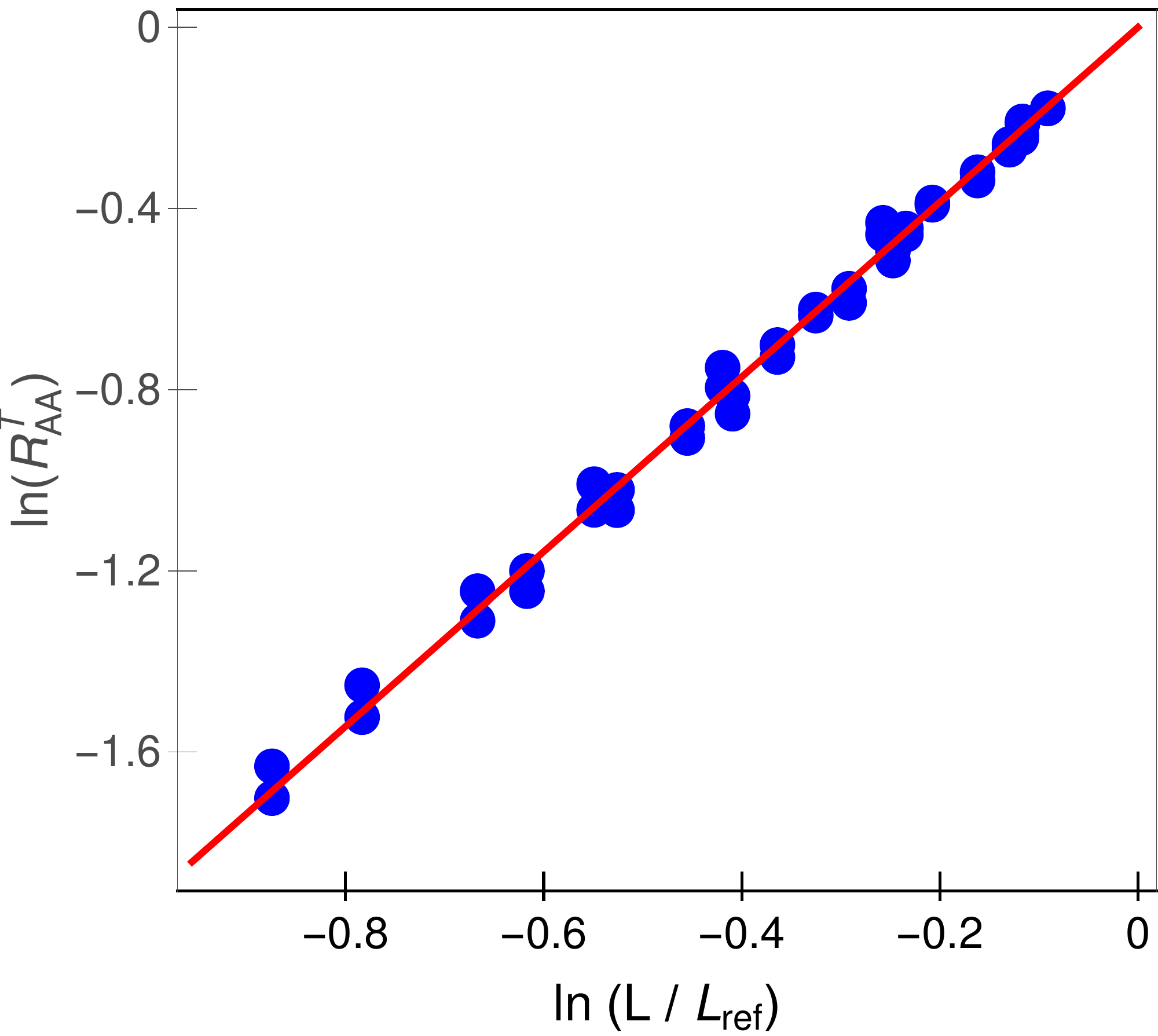}
  \caption{{\bf{Testing the validity of our procedure for temperature dependence extraction in the case of expanding QCD medium.}} $\ln (R^T_{AA})$ {\it vs.} $\ln(L/L_{ref})$ for $h^{\pm}$ at $p_{\perp} = 100$ GeV for different pairs of centrality classes is plotted. Suppression predictions are obtained from full-fledged DREENA-B~\citep{DREENAB} calculations. Referent centrality values are: ($ 5-10 \%, 10-20 \%, 20-30  \%, 30-40  \%, 40-50  \%, 50-60 \%$), while their counterpart values are always higher, with highest being $60-70 \%$. The red solid line corresponds to the linear fit to the values.}
  \label{fig:sl8}
\end{figure}
we may write:
\begin{equation}
    \frac{L}{L_{ref}} = \bigg( \frac{T_0}{T_{0,ref}} \bigg)^{\kappa} \implies \frac{T_0}{T_{0,ref}} = \bigg( \frac{L}{L_{ref}} \bigg) ^{1/\kappa},
    \label{LTkappa}
\end{equation}
which ensures that the $R^T_{AA}$  quantity has a very simple form, depending only on average path-length and exponents $a, b$ and $\kappa$:
\begin{equation}
    \begin{split}
R^T_{AA} &= \bigg( \frac{L}{L_{ref}} \bigg) ^ {\frac{a}{\kappa}+b-\frac{a}{3}}. \\
    \end{split}
    \label{RatioBjorken2}
\end{equation}

If we substitute value of $a \approx 1.2$ obtained in constant $T$ medium case, previously estimated $b \approx 1.4$~\citep{NewObserv} and here inferred $\kappa \approx 1.3$, we arrive at the following estimate:
\begin{equation}
 R^T_{AA} =   \bigg( \frac{L}{L_{ref}} \bigg) ^ {1.93} \implies
     \ln (R^T_{AA}) =1.93 \cdot \ln \bigg( \frac{L}{L_{ref}} \bigg).
     \label{RatioBjorkenFinal}
\end{equation}
This equation is quite suitable for testing the robustness of the procedure for extracting the exponent $a$ to inclusion of the evolving medium. Namely, value 1.93 in Eq.~\eqref{RatioBjorkenFinal} stems from coefficient $a$, which is extracted from constant $T$ medium case. On the other hand, if we plot $\ln(R^T_{AA})$, generated by full-fledged DREENA-B calculations (i.e., in evolving medium)  which is {\it fundamentally different} from DREENA-C,  against $\ln(L/L_{ref})$ for variety of centrality pairs,  again we observe a linear dependence (see Fig.~\ref{fig:sl8}). Furthermore, a linear fit to the values surprisingly  yields the exact same slope coefficient value  of 1.93.

Consequently, the procedure of extracting temperature dependence exponent, introduced first in the case of constant $T$ medium, is applicable to the expanding medium as well. The displayed consistency of the results provide confidence to general applicability  of the  procedure presented in this paper (suggesting robustness to the applied model of bulk medium) and supports the reliability of the value of extracted $T$ dependence exponent $a \approx 1.2$.

\subsection{Effects of colliding system size}

We below extend our analysis to smaller colliding systems in order to assess generality of the conclusions presented above. Smaller colliding systems, such as Xe + Xe, Kr + Kr, Ar + Ar and O + O are important to gradually resolve the issue of QGP formation in small systems (such as pA), and (except Xe + Xe, which is already in a run) are expected to be a part of the future heavy-ion program at the LHC~\citep{futureExp}.

As already discussed in~\citep{NewObserv}, for this analysis within DREENA-C framework~\citep{DREENAC} (which we employ here for simplicity, since the robustness of the procedure to evolving medium was demonstrated above) note that $R_{AA}$ depends on: {\it i)} initial high-$p_\perp$ parton distribution, {\it ii)} medium average ${T}$ and {\it iii)} path-length distribution. For different colliding systems (probably at slightly different $\sqrt{s_{NN}} = 5.44$ TeV compared to Pb + Pb system) we employ the same high-$p_\perp$ distributions, since in~\citep{b} it was shown that for almost twofold increase of the collision energy (from 2.76 TeV to 5.02 TeV)  the change in corresponding initial distributions results in a negligible change (approximately $5\%$) in suppression.

 Regarding the average temperature, one should note that $T$  is directly proportional to the charged particle multiplicity, while inversely proportional to the size of the overlap area and average medium size~\citep{NonC,DREENAC,NewObserv,SHEME}, i.e., ${T}\propto(\frac{dN_{ch}/d \eta}{A_\perp {L}})^{1/3}$. The transition to smaller
\begin{figure}
  \includegraphics[width=0.4\linewidth]{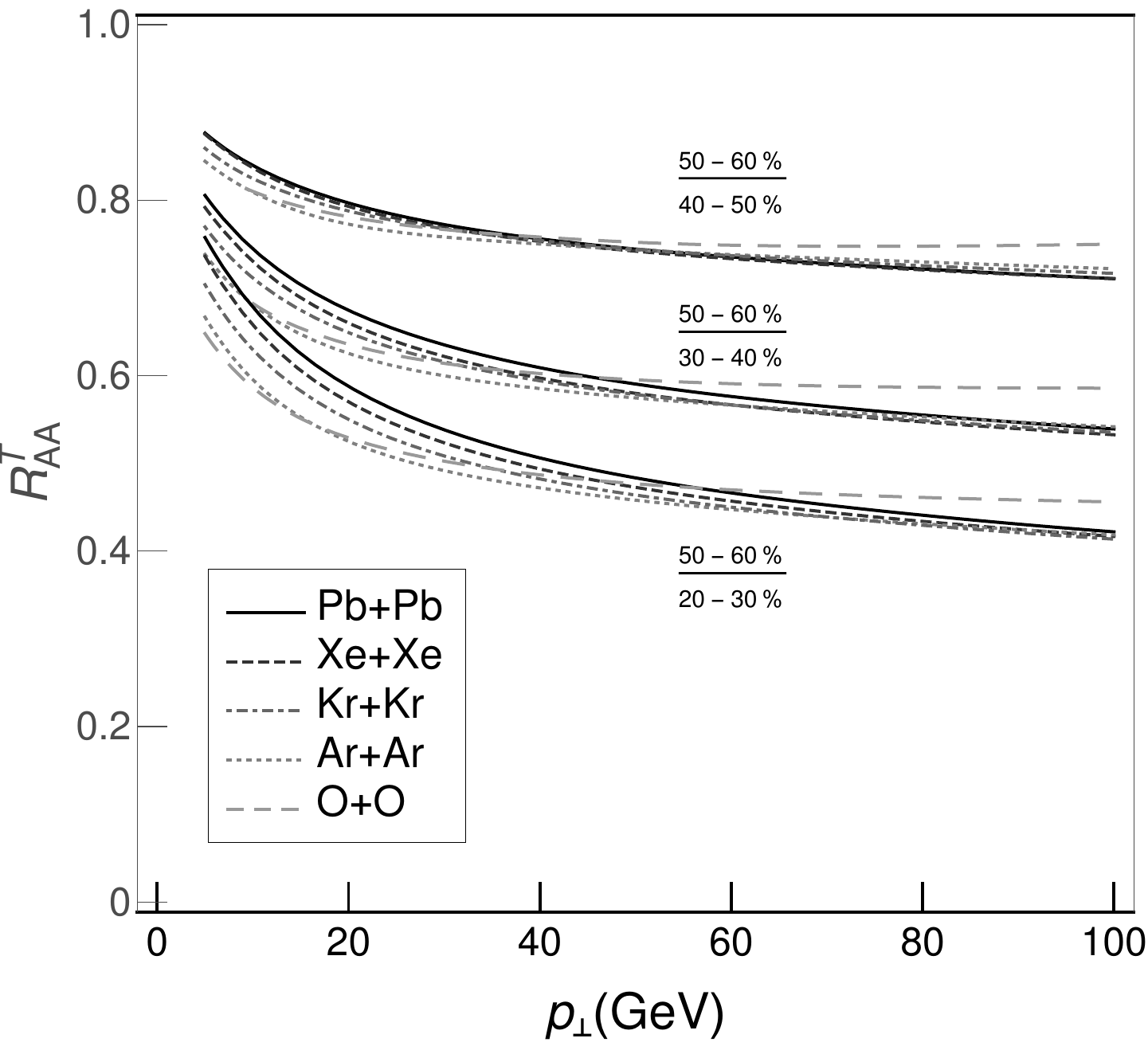}
  \caption{{\bf{Dependence of $\boldsymbol{R^T_{AA}}$ on a system size as a function of $\boldsymbol{p_{\perp}}$.}} Predictions for $h^\pm$ generated within full-fledged DREENA-C~\citep{DREENAC} suppression numerical procedure are compared for different colliding systems: Pb+Pb, Xe + Xe, Kr + Kr, Ar + Ar, O + O (for lines specification see legend). For clarity, the results are shown only for three centrality pairs, as specified in plot,  although checked for all available centrality classes. Magnetic to electric mass ratio is fixed to $\mu_M / \mu_E =0.4$.}
  \label{fig:sl5}
\end{figure}
colliding systems, for a certain fixed centrality class, leads to the following scaling: $A_\perp \propto A^{2/3}$, ${L} \propto A^{1/3}$~\citep{LA,LA1} and $dN_{ch}/ d \eta \propto N_{part} \propto A$~\citep{NA,NA1}, where $A$ denotes atomic mass. This leads to ${T} \sim (\frac{A}{A^{\frac{2}{3}} A^{\frac{1}{3}}})^{1/3} \sim const$, that is, we expect that average temperature does not change, when transitioning from large Pb + Pb to smaller systems, for a fixed centrality class. Lastly, path-length distributions for smaller systems and each centrality class are obtained in the same manner as for Pb+Pb~\citep{DREENAC}, and are the same as in Pb + Pb collisions up to a rescaling factor of $A^{1/3}$.

By denoting all quantities related to smaller systems with a tilde, with Pb + Pb quantities denoted as before, it is straightforward to show that the temperature sensitive suppression ratio for smaller systems satisfies:
\begin{align}
 \widetilde{R}^T_{AA} = \frac{1-\widetilde{R}_{AA}}{1-\widetilde{R}^{ref}_{AA}} &\approx \frac{\widetilde{{T}}^a \widetilde{{L}}^b}{ \widetilde{{T}}_{ref}^a \widetilde{{L}}_{ref}^b }  \approx  \frac{{T}^a {L}^b}{{T}_{ref}^a {L}_{ref}^b } \cdot \frac{ (\widetilde{A}/A)^{b/3} }{ (\widetilde{A}/A)^{b/3} } = \frac{1-R_{AA}}{1-R^{ref}_{AA}} = R^T_{AA},
    \label{small_sys_eq}
\end{align}
where we used: $\widetilde{{T}} = {T}$ and $\widetilde{{L}} /{L} = (\widetilde{{A}} / {A})^{1/3}$.

To validate equality of $R^T_{AA}$s for different system sizes, predicted by analytical scaling behavior (Eq.~\eqref{small_sys_eq}), in Fig.~\ref{fig:sl5} we compare our full-fledged $R^T_{AA}$ predictions for $h^\pm$ in Pb + Pb system with those for smaller colliding systems. We observe that, practically irrespective of system size,  $R^T_{AA}$ exhibits the same asymptotical behavior at high-$p_\perp$. This not only validates our scaling arguments, but also demonstrates the robustness of the new observable $R^T_{AA}$ to system size. Consequently,  since for fixed centrality range, ${T}$ should remain the same for all these colliding systems, we obtained that temperature dependence exponent $a$ should be the same independently of considered colliding system (see Fig.~\ref{fig:sl4}).  Therefore, the proposed procedure for extracting the temperature dependence of the energy loss is also robust to the collision system size. As a small exception, O + O system exhibits slight departure from the remaining systems at high-$p_\perp$, which might be a consequence of the fact that this system is significantly smaller than other systems considered here.

\section{Conclusions and outlook}

One of the main signatures of high-$p_\perp$ particle's energy loss, apart from its path-length, is its temperature dependence. Although extensive studies on both issues were performed, not until recently the path dependence resolution was suggested~\citep{NewObserv}. Here we proposed a new simple observable for extracting temperature dependence of the energy loss, based on one of the most common jet quenching observable $-$ the high-$p_\perp$ suppression. By combining full-fledged numerical calculations with asymptotic scaling behavior, we surprisingly obtained that temperature dependence is nearly linear, i.e., far from quadratic or cubic, as commonly assumed. Further, we verified its robustness and reliability on colliding system size and evolving QGP medium. Moreover, we demonstrated that the same observable can be utilized to discriminate between different energy loss models on {\it both} their {\it temperature} and {\it path length} dependence bases. Comparison with the experimental data also indicated a need for revising the long-standing $\Delta E/E \propto L^2 T^3$ paradigm.

As an outlook, the expected substantial decrease of error-bars in the upcoming $3^{rd}$ run measurements at the LHC  will allow direct extraction of temperature dependence exponent from high-$p_\perp$ data of this observable. This will provide a resolving power to temperature/path-length~\citep{NewObserv} dependence of the energy loss  and test our understanding of the underlying QGP physics.




{\em Acknowledgments:}
We thank Pasi Huovinen and Jussi Auvinen for useful discussions. This work is supported by the European Research Council, grant ERC-2016-COG: 725741, and by the Ministry of Science and Technological
Development of the Republic of Serbia, under project No.  ON171004 and ON173052.

\end{document}